\documentclass[twocolumn,prb]{revtex4}
\usepackage{amsfonts}
\usepackage[T1]{fontenc}
\usepackage{amsmath,amsbsy,amssymb,graphicx}
\usepackage{times}

\begin{document}

\title{Quench dynamics and bulk-edge correspondence in nonlinear mechanical
systems}
\author{Motohiko Ezawa}
\affiliation{Department of Applied Physics, University of Tokyo, Hongo 7-3-1, 113-8656,
Japan}

\begin{abstract}
We study a topological physics in a one-dimensional nonlinear system by
taking an instance of a mechanical rotator model with alternating spring
constants. This nonlinear model is smoothly connected to an acoustic model
described by the Su-Schrieffer-Heeger model in the linear limit. We
numerically show that quench dynamics of the kinetic and potential energies
for the nonlinear model is well understood in terms of the topological and
trivial phases defined in the associated linearized model. It indicates
phenomenologically the emergence of the edge state in the topological phase
even for the nonlinear system, which may be the bulk-edge correspondence in
nonlinear system.
\end{abstract}

\maketitle

\section{Introduction}

Topological insulators\cite{Hasan,Qi} were originally discovered in
in-organic solid state materials. However, it is now well recognized that
topological physics is ubiquitous in various systems such as acoustic\cite%
{Prodan,TopoAco,Berto,Xiao,He,Abba,Xue,Ni,Wei,Xue2}, mechanical\cite%
{Lubensky,Chen,Nash,Paul,Sus,Sss,Huber,Mee,Kariyado,Hannay,Po,Rock,Takahashi,Mat,Taka,Ghatak,Wakao}%
, photonic\cite{KhaniPhoto,Hafe2,Hafezi,WuHu,TopoPhoto,Ozawa,Hassan,Li} and
electric circuit\cite%
{TECNature,ComPhys,Hel,Lu,YLi,EzawaTEC,Research,Zhao,EzawaLCR,EzawaSkin,Garcia,Hofmann,EzawaMajo,Tjunc,Lee,Kot}
systems. They are called artificial topological systems. The merit of them
is that it is possible to fabricate ideal systems comparing to natural solid
state materials. Another merit of artificial topological systems is that
nonlinearity is naturally introduced into them.

Topological physics is mostly studied in linear systems. There are only a
few studies on it in nonlinear systems\cite{Kot,Smi,Sone,TopoToda} because
the study of the topological properties is not straightforward. One of the
reasons is that it is a formidable problem to obtain a band structure and a
topological number in a generic nonlinear system. Recently, it is proposed
to account for the topological properties in nonlinear systems
phenomenologically based on the bulk-edge correspondence well established in
the linear theory\cite{TopoToda}. It seems to work provided that the
nonlinear theory is continuously connected to a linear theory where the
topological number is well defined. One may say that the topological
properties are inherited from a linear theory to a nonlinear theory. It is
an interesting problem to explore other systems which share similar
properties.

A good signal to detect the topological phase transition is to study quench
dynamics starting from one of the edges in the case of one dimension based
on the bulk-edge correspondence\cite{QWalk}. There is almost no time evolution
and the state remains almost localized at the edge for a topological phase.
It is because the initial state is almost given by the topological localized
edge state, which has no dynamics. On the other hand, the state rapidly
spreads into the bulk for a trivial phase because the initial state is
composed of bulk eigen functions. Although the usefulness of quench dynamics
has been established in a linear system, it is also useful in a nonlinear
system\cite{TopoToda}.

\begin{figure}[t]
\centerline{\includegraphics[width=0.48\textwidth]{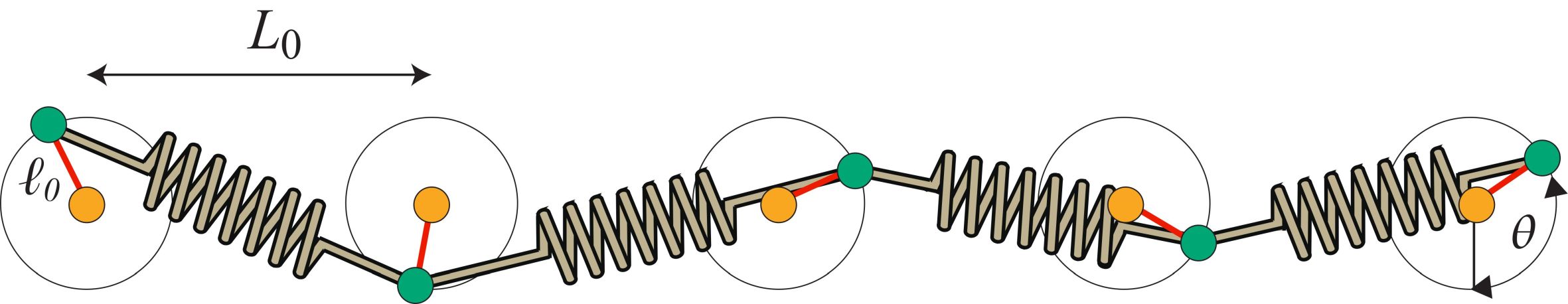}}
\caption{Illustration of the mechanical rotator model. Two adjacent rotators
are connected by a spring with the spring constant $\protect\kappa _{j}$. A
rotator colored in green rotates around a center colored in orange with the
radius $\ell _{0}$. All centers are on a line, and the distance between two
adjacent centers is $L_{0}$. }
\label{FigRing}
\end{figure}

In this paper, we investigate a one-dimensional mechanical rotator model
with alternating spring constants: See Fig.\ref{FigRing}. It is a nonlinear
generalization of the Su-Schrieffer-Heeger (SSH) model. We solve the quench
dynamics of a mechanical rotator model as an initial condition problem,
where only one rotator at the left-most edge is excited initially. The
potential and kinetic energies exhibit distinct behaviors depending on
whether the system is in the topological or trivial phase defined in the
linearized model. Namely, after enough time, they are well localized in the
topological phase, while they are spread over the bulk in the trivial phase.
These phenomena are understood in terms of the emergence of the topological
edge state in the topological phase. It would represent the bulk-edge
correspondence in nonlinear system.

\section{Mechanical rotator model}

We consider a mechanical rotator model\cite{Lubensky} as illustrated in Fig.%
\ref{FigRing}. We place rotators indexed by the number $j$, $j=1,2,\cdots ,N$%
. Each rotator has the radius $\ell _{0}$, and its center is fixed at the
position $\left( jL_{0},0\right) $ on the $x$ axis, with $L_{0}$ the
distance between two adjacent centers. The dynamical variables are angles $%
\theta _{j}$ for rotators $j$.

The Hamiltonian of the system consists of the kinetic energy $U^{\text{K}}$,
the potential energy of the springs $U^{\text{spring}}$ and the
gravitational energy of the $U^{\text{g}}$,%
\begin{equation}
H=U^{\text{K}}+U^{\text{spring}}+U^{\text{g}}.
\end{equation}%
They are given by 
\begin{equation}
U^{\text{K}}=\sum_{j}U_{j}^{\text{K}},\quad U_{j}^{\text{K}}=\frac{m}{2}\dot{%
\theta}_{j}^{2}
\end{equation}%
with the mass $m$, 
\begin{equation}
U^{\text{g}}=-g\sum_{j}\cos \theta _{j}
\end{equation}%
with the gravitational constant $g$, and 
\begin{eqnarray}
U^{\text{spring}} &=&\sum_{j}U_{j}^{\text{spring}}, \\
U_{j}^{\text{spring}} &=&\frac{1}{2}\kappa _{j}\left( L_{j}\left( \theta
_{j},\theta _{j+1}\right) -L\right) ^{2},
\end{eqnarray}%
where $L_{j}\left( \theta _{j},\theta _{j+1}\right) $ is the length of the
spring between $j$ and $j+1$ nodes,%
\begin{equation}
L_{j}\left( \theta _{j},\theta _{j+1}\right) =\sqrt{\left[ L_{j}^{x}\left(
\theta _{j},\theta _{j+1}\right) \right] ^{2}+\left[ L_{j}^{y}\left( \theta
_{j},\theta _{j+1}\right) \right] ^{2}}
\end{equation}%
with

\begin{figure}[t]
\centerline{\includegraphics[width=0.48\textwidth]{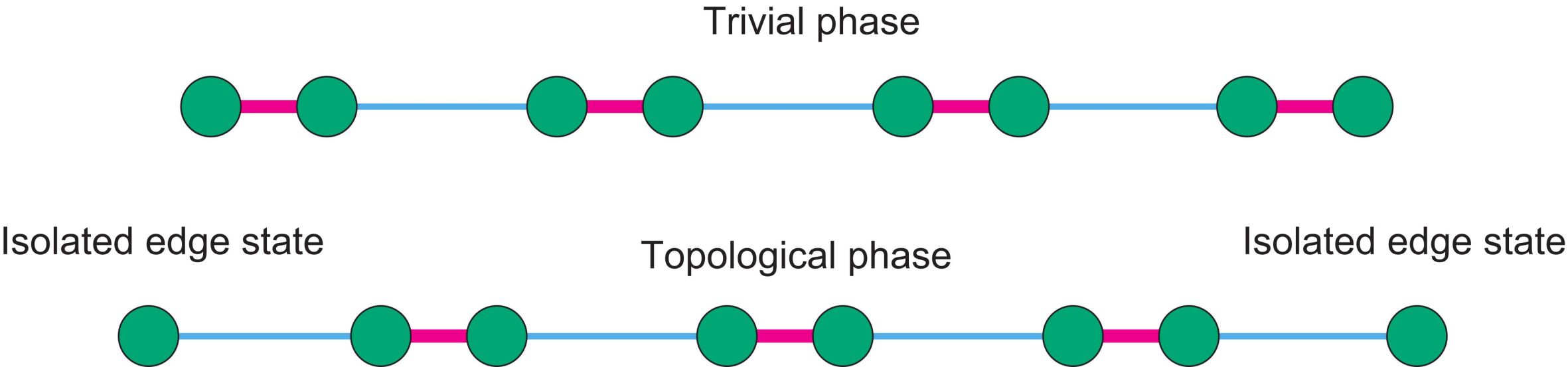}}
\caption{Illustration of the trivial and topological phases in the SSH
model. There are two isolated edge states in the topological phase, while
all of the sites are dimerized in the trivial phase. Magenta and cyan lines
represent strong and weak bondings $\protect\kappa _{j}$ in Eq.(\protect\ref%
{SpringCon}), respectively. This illustration is argued to hold even in
nonlinear system described by Eq.(\protect\ref{NLEq}). }
\label{FigSSH}
\end{figure}

\begin{eqnarray}
L_{j}^{x}\left( \theta _{j},\theta _{j+1}\right) &=&\ell _{0}\sin \theta
_{j+1}-\ell _{0}\sin \theta _{j}+L_{0}, \\
L_{j}^{y}\left( \theta _{j},\theta _{j+1}\right) &=&\ell _{0}\cos \theta
_{j+1}-\ell _{0}\cos \theta _{j}.
\end{eqnarray}%
The equation of motion is given by%
\begin{equation}
m\ddot{\theta}_{j}=-g\sin \theta _{j}-\frac{\partial U^{\text{spring}}}{%
\partial \theta _{j}}.  \label{NLEq}
\end{equation}%
The spring constant is assumed to be alternating,%
\begin{equation}
\kappa _{j}=\kappa \left( 1+\lambda \left( -1\right) ^{j}\right) ,
\label{SpringCon}
\end{equation}%
where the dimerization is controlled by $\lambda $ with $|\lambda |\leq 1$.
For $\lambda <0$, the spring constant $\kappa _{j}$ with odd (even) $j$ is
strong (weak). On the other hand, for $\lambda >0$, the spring constant $%
\kappa _{j}$ with odd (even) $j$ is weak (strong): See Fig.\ref{FigSSH}.

\begin{figure*}[t]
\centerline{\includegraphics[width=0.98\textwidth]{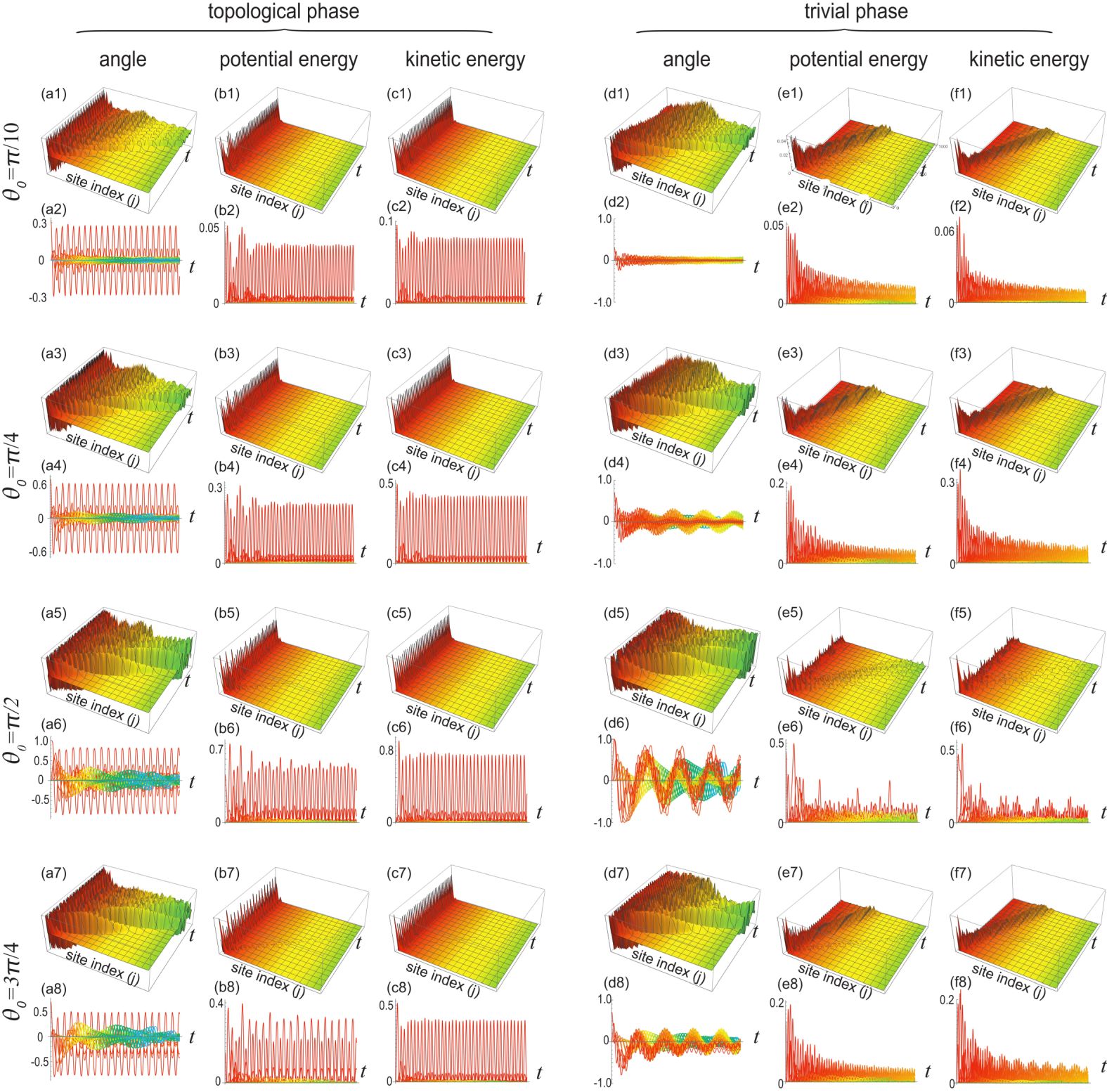}}
\caption{Time evolution of the angle $\sin \protect\theta _{j}$ for (a1)$%
\sim $(a8) and (d1)$\sim $(d8), the potential energy $U_{j}^{\text{spring}}$
for (b1)$\sim $(b8) and (e1)$\sim $(e8), and the kinetic energy $U_{j}^{%
\text{K}}$ for (c1)$\sim $(c8) and (f1)$\sim $(f8). (a1)$\sim $(c8)
topological phase with $\protect\lambda =-0.5$, and (d1)$\sim $(f8) trivial
phase with $\protect\lambda =0.5$. (a1)$\sim $(f2) $\protect\theta _{0}=%
\protect\pi /10$, (a3)$\sim $(f4) $\protect\theta _{0}=\protect\pi /4$, (a5)$%
\sim $(f6) $\protect\theta _{0}=\protect\pi /2$, and (a7)$\sim $(f8) $%
\protect\theta _{0}=3\protect\pi /4$. (a1)$\sim $(f1), (a3)$\sim $(f3), (a5)$%
\sim $(f5) and (a7)$\sim $(f7) Bird's eye's view, and (a2)$\sim $(f2), (a4)$%
\sim $(f4), (a6)$\sim $(f6) and (a8)$\sim $(f8) Color plot of the time
evolution. We have used a chain containing 100 sites. We have set $m=0.01$, $%
g=0.1$, $L_{0}=10$ and $\ell _{0}=1$. Color indicates the site index $j$,
which is manifest in bird's eye's views.}
\label{FigDynamics}
\end{figure*}

\section{Linearized model}

Provided the angle $\theta _{j}$ is small enough, the potential energy is
well approximated by the harmonic potential,%
\begin{equation}
U^{\text{spring}}=\frac{\ell _{0}^{2}}{2}\sum_{j}\kappa _{j}\left( \theta
_{j+1}-\theta _{j}\right) ^{2},
\end{equation}%
and the equation of motion is obtained as%
\begin{equation}
m\ddot{\theta}_{j}=-g\theta _{j}-\ell _{0}^{2}\sum_{j}\kappa _{j}\left(
\theta _{j}-\theta _{j+1}\right) +\kappa _{j-1}\left( \theta _{j}-\theta
_{j-1}\right) ,  \label{LEq}
\end{equation}%
which is a linearized model.

\begin{figure*}[t]
\centerline{\includegraphics[width=0.98\textwidth]{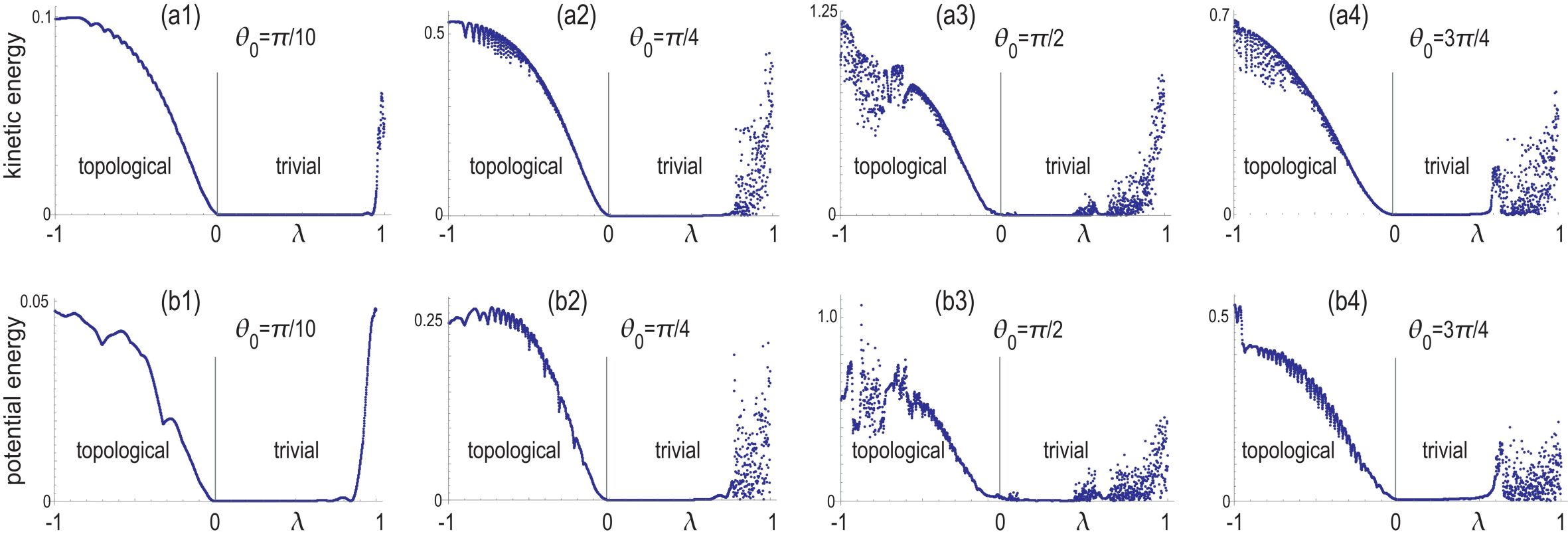}}
\caption{(a1)$\sim $(a4) Kinetic energy at the edge $U_{1}^{\text{K}}$, and
(b1)$\sim $(b4) the potential energy at the edge $U_{1}^{\text{spring}}$ as
a function of the initial position $\protect\theta _{0}$ after enough time.
(a1) and (b1) $\protect\theta _{0}=\protect\pi /10$, (a2) and (b2) $\protect%
\theta _{0}=\protect\pi /4$, (a3) and (b3) $\protect\theta _{0}=\protect\pi %
/2$, (a4) and (b4) $\protect\theta _{0}=3\protect\pi /4$. See the caption of
Fig.\protect\ref{FigDynamics} for the parameters. }
\label{FigDistribution}
\end{figure*}

Eq.(\ref{LEq}) is rewritten in the form of%
\begin{equation}
m\ddot{\theta}_{j}=\sum_{j}K_{ij}\theta _{j},
\end{equation}%
where 
\begin{equation}
K_{ij}=-\delta _{ij}\left( g+\ell ^{2}\right) +\ell _{0}^{2}\left( \kappa
_{j}\delta _{i,j+1}+\kappa _{j-1}\delta _{i,j-1}\right)
\end{equation}%
is identical to the SSH Hamiltonian. After a Fourier transformation, we have%
\begin{equation}
K\left( k\right) =-\left( g+\ell _{0}^{2}\right) I_{2}+\kappa \left( 
\begin{array}{cc}
0 & 1-\lambda e^{-ik} \\ 
1-\lambda e^{ik} & 0%
\end{array}%
\right) .  \label{EqK}
\end{equation}%
It is known\cite{Lubensky,Chen} that the system is topological for $\lambda
<0$ and trivial for $\lambda >0$. There are two isolated edge states in the
limit $\lambda \simeq -1$, while all of the states are dimerized in the
limit $\lambda \simeq 1$: See Fig.\ref{FigSSH}.

The topological number associated with the SSH Hamiltonian is the chiral
index defined by%
\begin{equation}
\Gamma =\int_{0}^{2\pi }\text{Tr}\left[ \sigma _{z}K^{-1}\partial _{k}K%
\right] dk,  \label{ChiralIndex}
\end{equation}%
where $K$ is given by Eq.(\ref{EqK}). We obtain $\Gamma =1$ for $\lambda <0$
and $\Gamma =0$ for $\lambda >0$.

Our interest is the effect of the nonlinearity based on Eq.(\ref{NLEq})
instead of the linear equation of motion (\ref{LEq}).

\section{Nonlinear quench dynamics and bulk-edge correspondence}

The quench dynamics starting from a state localized at one edge well
captures the topological phase transition in linear systems\cite{QWalk},
where the eigenfunctions are easily obtained and a topological number is
well defined. In the topological phase, there are topological edge states
well localized at edges, which is known as the bulk-edge correspondence. If we
excite only one edge site, the most component is dominated by a topological
edge state. The topological edge state remains as it is after time
evolution. On the other hand, there is no such localized edge state in the
trivial phase. Hence, all of the components of the initial state are bulk
eigenfunctions. They are spread into the bulk after the time evolution.
Thus, it is possible to distinguish topological and trivial phases by
checking whether the state is localized or spread into the bulk.

The above observation is also applicable even for nonlinear systems\cite%
{TopoToda}. The existence of the localized topological edge state is obscure
in nonlinear systems because it is not possible to diagonalize the
Hamiltonian and obtain eigenfunctions. Nevertheless, the quench dynamics
shows distinct behaviors between the topological and trivial phases as in
the case of the linear system. Thus, the quench dynamics starting from one
of the edges is a strong signal to detect a topological phase transition.

We numerically solve the equation of motion (\ref{NLEq}) with the initial
condition $\theta _{j}\left( t=0\right) =\theta _{0}\delta _{j,1}$ and $\dot{%
\theta}_{j}\left( t=0\right) =0$. If $\left\vert \theta _{0}\right\vert \ll
1 $, the system is well described by the linear equation (\ref{LEq}).
Otherwise, the system is nonlinear. We study a case $\theta _{0}=\pi /10,\pi
/4,\pi /2$ and $3\pi /4$ as typical examples.

We show the time evolution of $\sin \theta _{j}$, the potential energy $%
U_{j}^{\text{spring}}$ and the kinetic energy $U_{j}^{\text{K}}$ in Fig.\ref%
{FigDynamics}. The quench dynamics of $\sin \theta _{j}$ is significantly
different between the topological and trivial phases. The amplitude of $\sin
\theta _{1}$ at the left edge site remains finite in the topological phase.
On the other hand, the amplitude of $\sin \theta _{1}$ decreases in the
trivial phase. The potential energy $U_{j}^{\text{spring}}$ remains well
localized at the left edge in the topological phase although the absolute
value of $\sin \theta _{j}$ spreads into the bulk as shown in Fig.\ref%
{FigDynamics}(b). On the other hand, it moves as if it were a soliton in the
trivial phase as shown in Fig.\ref{FigDynamics}(e). If $\theta _{0}=\pi /2$,
there remains a finite component in $U_{j}^{\text{spring}}$ as well as a
soliton-like propagation. This will be because that $\sin \theta _{j}$ takes
a maximum for $\theta _{0}=\pi /2$, where the dynamics is most enhanced and
it takes longer time to reach a steady state.

We also show the time evolution of the kinetic energy $U_{j}^{\text{K}}$ in
Fig.\ref{FigDynamics}(c) and (f). The behavior of $U_{j}^{\text{K}}$ is
quite similar to that of $U_{j}^{\text{spring}}$.

The dynamics between $\theta _{0}=\pi /4$ and $3\pi /4$ are very similar as
shown in Fig.\ref{FigDynamics}. It will be due to the fact that $\sin \pi
/4=\sin 3\pi /4$ although we have $\cos \pi /4\neq \cos 3\pi /4$. These
results indicate a symmetry between the angle $\theta _{1}$ and $\theta _{2}$
satisfying the condition $\sin \theta _{1}=\sin \theta _{2}$, where the
dynamics is similar.

The potential and kinetic energies after enough time present a good signal
for the topological phase transition comparing to the dynamics of $\sin
\theta _{j}$. We show the kinetic energy $U_{j}^{\text{K}}$ and the
potential energy $U_{j}^{\text{spring}}$\ after enough time as a function of 
$\lambda $ in Fig.\ref{FigDistribution}. It is finite for the topological
phase with $\lambda <0$, while it is almost zero for the trivial phase with $%
\lambda >0$. These features are common for all initial conditions $\theta
_{0}=\pi /10,\pi /4,\pi /2$ and $3\pi /4$. It indicates the bulk-edge
correspondence in nonlinear systems. It suggests the validity of the
topological number (\ref{ChiralIndex}) even in strong nonlinear regime.

A comment is in order with respect to finite components around $\lambda
\simeq 1$\ in Fig.\ref{FigDistribution}. They are interpreted as follows. In
the limit $\lambda \simeq 1$, the system is almost dimerized as shown in Fig.%
\ref{FigSSH}, where there are no isolated sites at the both edges. In this
limit, the energy is not well transferred to the bulk because the energy is
localized in the dimer located at the edge to which the energy is injected,
resulting in finite components around $\lambda \simeq 1$.

\section{Discussion}

We have shown that the topological and trivial phases are well
differentiated in the mechanical rotator model even in a strong nonlinear
regime based on the bulk-edge correspondence. Our finding is that the
topological properties are inherited to the nonlinear model from the
associated linearized model provided they are smoothly connected. Then, it
would be possible to use a topological number defined in the linearized
model. This phenomenon is quite similar to the one in the Toda lattice\cite%
{TopoToda}, which is a typical exactly solvable model containing a soliton.

The author is very much grateful to N. Nagaosa for helpful discussions on
the subject. This work is supported by the Grants-in-Aid for Scientific
Research from MEXT KAKENHI (Grants No. JP17K05490 and No. JP18H03676). This
work is also supported by CREST, JST (JPMJCR16F1 and JPMJCR20T2).

\end{document}